\documentclass[pra,showpacs,superscriptaddress,amsfonts,amsmath,floatfix,twocolumn]{revtex4-1}
\usepackage{graphicx}
\usepackage{hyperref} 
\usepackage{verbatim} 
\usepackage{epstopdf}
\usepackage[normalem]{ulem}
\usepackage{natbib}
\usepackage{color}
\usepackage{dsfont}
\usepackage[utf8]{inputenc}
\usepackage{float}
\usepackage{amsmath}
\usepackage{comment}
\usepackage{breqn}
\usepackage{amssymb}
\usepackage[font=footnotesize,labelfont=bf]{caption}
\renewcommand\vec[1]{\ensuremath\boldsymbol{#1}}
\newcommand{\affA}{Department of Physics and Astronomy, Aarhus University, Ny Munkegade 120, Denmark}
\newcommand{\affB}{Instituto de F{\'i}sica da UFRGS, Av. Bento Gon{\c c}alves 9500, Porto Alegre, RS, Brazil}

\begin{document}
\title{Dynamical realization of magnetic states in a strongly interacting Bose mixture}

\author{R. E. Barfknecht}
\email{rafael.barfknecht@ufrgs.br}
\affiliation{\affB}
\affiliation{\affA}

\author{A. Foerster}
\email{angela@if.ufrgs.br}
\affiliation{\affB}

\author{N. T. Zinner}
\email{zinner@phys.au.dk}
\affiliation{\affA}

\date{\today}
\begin{abstract}
\noindent We describe the dynamical preparation of magnetic states in a strongly interacting two-component Bose gas in a harmonic trap. By mapping this system to an effective spin chain model, we obtain the dynamical spin densities and the fidelities for a few-body system. We show that the spatial profiles transit between ferromagnetic and antiferromagnetic states as the intraspecies interaction parameter is slowly increased.

\end{abstract}


\maketitle

\section{Introduction}
The recent progress in magneto-optical trapping of ultracold atoms \cite{bloch1} has opened up a new area of experimental development in physics, allowing for the construction of paradigmatic models of quantum mechanics. One of the most important product of these advances is the realization of effective one-dimensional (1D) atomic systems \cite{moritz,weiss1,haller1,haller2} where interactions can be tuned via Feshbach \cite{pethick} or confinement induced resonances \cite{olshanii,haller3}. Specially relevant among these 1D experiments is the strongly repulsive bosonic system known as the Tonks-Girardeau gas \cite{paredes,weiss2}. The refinements in manipulation and controlling of cold atoms also enabled the probing of fundamental properties of quantum systems through the construction of few-body ensembles \cite{jochim1,jochim2,jochim3}.

From a theoretical standpoint, the problem of few particles interacting in a harmonic trap has been addressed through different approaches, both exact and approximative \cite{brouzos,rubeni1,rubeni2,artem1,bruno,amin1,barfk1}. The case of strongly interacting atoms, in particular, has been shown to be analogous to an effective 1D spin chain \cite{deuretz1,artem2,pu1,xiaoling}. Moreover, strongly interacting few-body systems are suitable for studying the origins of quantum magnetism \cite{deuretz2,amin2,massign} even in models without underlying lattices. Recently, it has been shown that different magnetic orderings can also be induced by adding p-wave interactions to the system \cite{xiaoling2}. While many of these works deal with static properties, the studies involving dynamical features such as spin transport \cite{fukuhara}, state transfer \cite{artem3,loft1} and time evolution following a sudden quench \cite{pu2} are less numerous. Nonetheless, they are of great experimental interest \cite{fukuhara}, specially due to their possible applications in spintronics, quantum information processing and communication \cite{bose}. Therefore, a more detailed investigation of quantum dynamics and magnetism in this few-body strongly interacting context is welcome and constitutes the main focus of this work.

Given the motivations above and viewing the possibility of experiments with ultracold few-body Bose mixtures, we consider a model of strongly interacting two-component bosonic atoms in a harmonic trap. It is known that different magnetic states arise as the interactions between bosonic or fermionic atoms are manipulated \cite{massign,haiping}. Here, we specifically show that the spin densities of the system transit between states with clear ferromagnetic (FM) and antiferromagnetic (AFM) profiles as the intraspecies interaction is increased in time. This transition is visible not only in the dynamical fidelities, but also in the spatial distribution of the spins in the trap. 

The paper is organized as follows: in section \ref{hamilt} we present the Hamiltonian for the strongly interacting two-component bosonic system and the mapping to an effective spin chain model. By considering the solution of the system in the infinite repulsion limit, we calculate the ground state spatial densities and the trap-dependent geometric coefficients. The system is then considered to be completely described only by the solution of the spin chain Hamiltonian. We choose to initialize the system in an eigenstate where the intraspecies interaction is smaller than the interspecies interaction. In section \ref{dyn} we proceed to obtain the dynamics of the system: by changing the intraspecies interaction in time and solving the eigenvalue problem at each time step, we can obtain the time evolution of the spin densities. We show that, for increasing intraspecies repulsion, the system evolves from an initial FM state and asymptotically reaches an AFM profile. We demonstrate this by calculating the time evolution of spin densities for different imbalanced systems. In the balanced case, although the spin densities provide less information when compared to the imbalanced situation, the squared fidelities still show the transitions between FM and AFM states. In section \ref{conc} we present the conclusions and future work perspectives.

\section{Hamiltonian and mapping to an effective spin chain}\label{hamilt}

We consider a trapped 1D Bose gas with contact interactions and two different bosonic species ($\uparrow,\downarrow$). The total number of particles is $N=N_{\uparrow}+N_{\downarrow}$ where $N_{\uparrow}$ and $N_{\downarrow}$ are the numbers of particles of species $\uparrow$ and $\downarrow$, respectively. The $N$-body hamiltonian is given by
\begin{eqnarray}\label{hm2}
H=\sum_i^{N} H_0(x_i)+g\sum_{\uparrow\downarrow}\delta(x_i-x_j)+\\ \kappa g  \nonumber \sum_{\uparrow\uparrow} \delta(x_i-x_j)+\kappa g\sum_{\downarrow\downarrow}\delta(x_i-x_j),
\end{eqnarray}
where we assume $\hbar=m=1$ and
\begin{equation}\label{sphm}
H_0(x)=-\frac{1}{2}\frac{\partial^2}{\partial x^2}+V(x)
\end{equation}
is the single particle hamiltonian for a given potential $V(x)$ (for harmonic trapping, we have $V(x)=x^2/2$). The remaining terms of the hamiltonian account for the contact interactions between particles of different species (with strength parameter $g$) and of the same species (with strength parameter $\kappa g$). We consider the length, time and energy units to be $l=\sqrt{\hbar/m\omega}$, $\tau=1/\hbar \omega$ and $\hbar \omega$, respectively, where $\omega$ is the longitudinal harmonic confinement frequency \cite{olshanii}.

\begin{figure}[H]
\centering
\includegraphics[scale=0.3]{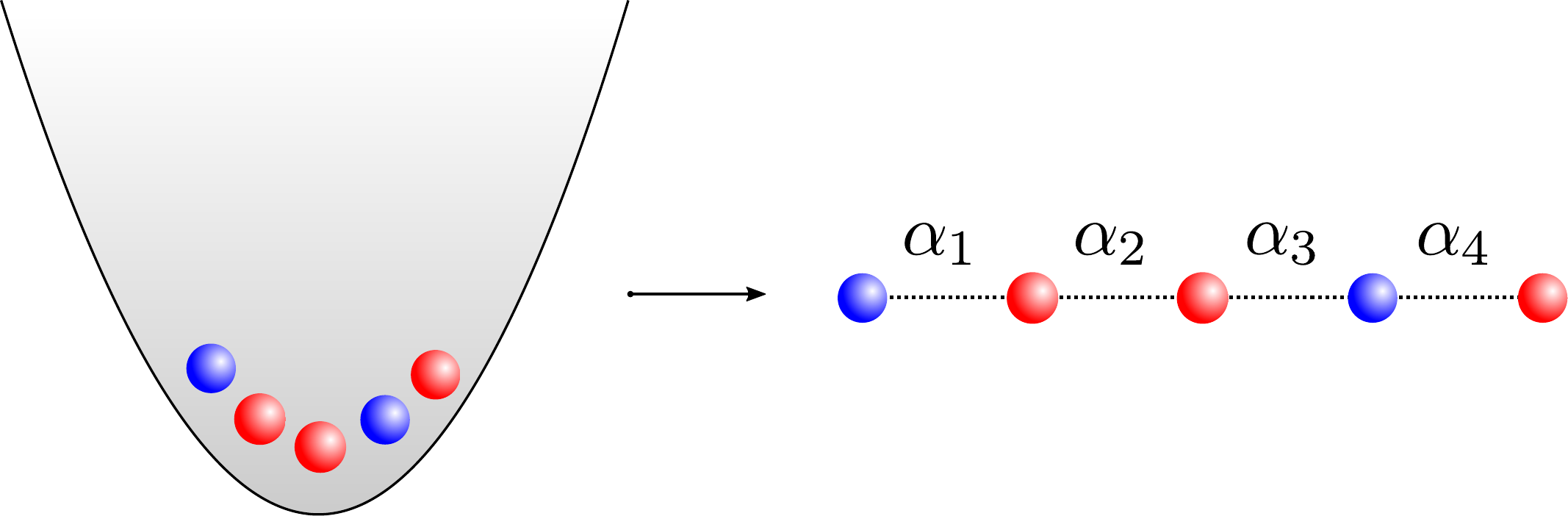}
\caption{(a) A system of strongly interacting atoms in a trapping potential can be mapped to an effective spin chain model where the coefficients $\alpha$ are determined by the geometry of the trap.}
\label{fig1}
\end{figure}

In the limit of infinite repulsion ($1/g=0$), the solution of this system is given by the Bose-Fermi mapping \cite{girardeau}. This wave function of hardcore bosons is a symmetrized Slater determinant constructed from the individual eigenstates of the single-particle Hamiltonian (\ref{sphm}). Its energy $E_0$ is simply the sum of the energies of the lowest occupied energy levels of the potential $V(x)$.

In the limit of strong interactions ($g\gg 1$), the Hamiltonian (\ref{hm2}) can be mapped, up to linear order in $1/g$, to a spin chain model given by
\begin{equation}\label{spinchain}
H_s=E_0-\sum_{i=1}^{N-1}\frac{\alpha_i}{g}\left[\frac{1}{2}(1-\vec{\sigma}^i\cdot\vec{\sigma}^{i+1})+\frac{1}{\kappa}(1+\sigma_z^{i}\sigma_z^{i+1})\right],
\end{equation}
where $\vec{\sigma}^{i}=(\sigma_x^i,\sigma_y^i,\sigma_z^i)$ are the Pauli matrices acting on site $i$ and $E_0$ is the energy of the hardcore boson (or spinless fermion) system. In the limit of $\kappa\rightarrow \infty$ and positive $g$ the identical bosons are non-interacting, while for $\kappa=1$, the interaction strength between all bosons is the same. In the particular case of $\kappa=2$ we have an effective $XX$ model, as summarized in Ref. \cite{artem2}. The spin model for bosons described in Ref. \cite{xiaoling} can be obtained from Eq.~\ref{spinchain} by performing a unitary transformation (see supplemental material of Ref. \cite{marchukov}).

The coefficients $\alpha$ depend only on the geometry of the trap and are obtained from \cite{artem2}
\begin{equation}\label{geo}
\alpha_i=\frac{\int_{x_1<x_2...<x_N-1}dx_1...dx_{N-1}\Big|\frac{\partial \Phi_0^2}{\partial x_N^2}\Big|^2_{x_N=x_i}}{\int_{x_1<x_2...<x_N-1}dx_1...dx_N |\Phi_0^2|},
\end{equation}
where  $\Phi_0(x_1,...x_N)$ is the wave function for spinless fermions. An efficient computational scheme for obtaining the $\alpha$'s as the number of atoms $N$ is increased is presented in Ref. \cite{conan}.

In Fig.~\ref{fig1} (a) we represent the mapping from a strongly interacting 1D system in a harmonic trap to a spin chain characterized by the geometric coefficients $\alpha$. We will mainly focus on the $N=5$ problem, for which we obtain $\alpha_1=\alpha_4=2.16612$ and $\alpha_2=\alpha_3=3.17738$ (since the trap is symmetric, we have that $\alpha_i=\alpha_{N-i}$). Due to a factor of $1/2$ in the spin chain Hamiltonian, our geometric coefficients $\alpha_i$ are twice as large as the ones calculated in Ref. \cite{massign}.

\subsection{One-body correlations for the hardcore boson system}
\begin{figure}[H]
\centering
\includegraphics[scale=0.4]{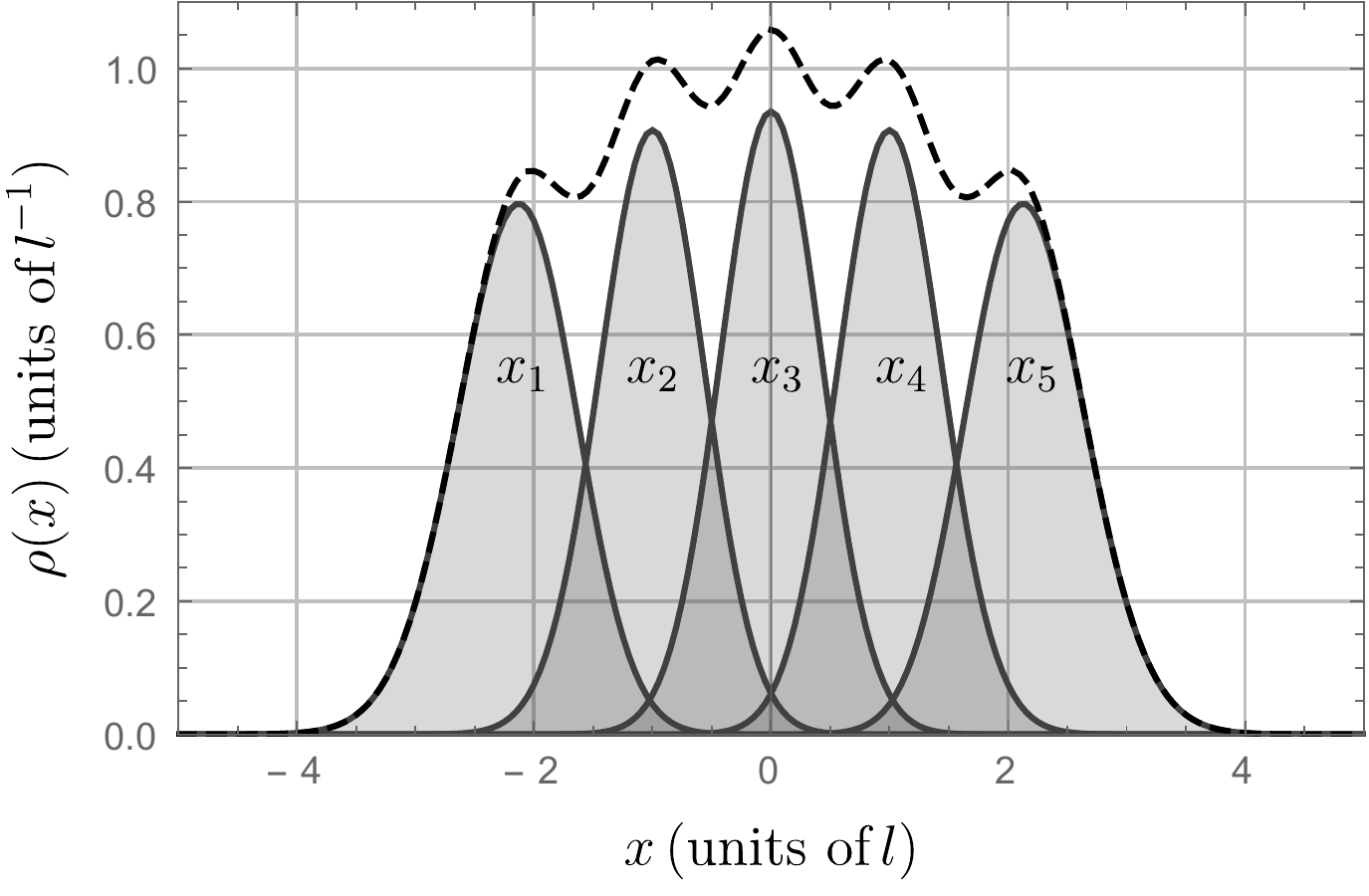}
\caption{One-body densities for $N=5$, calculated for the sector $x_1\leq...\leq x_i \leq...\leq x_5$. The total density (black dashed curve) is normalized to $N$.}
\label{fig2}
\end{figure}

We focus initially in obtaining the one-body densities for the hardcore boson system, since this accounts for the spatial part of the wave functions. The spatially ordered one-body correlations are given by
\begin{equation}\label{onebody}
\rho^i(x)=\int dx_1...dx_N \,\delta(x_i-x)|\Phi_0(x_1,...,x_i,...,x_N)|^2,
\end{equation}
where $\delta(x)$ is the Dirac delta function, and the integration is restricted to the sector $x_1\leq...\leq x_i \leq...\leq x_N$.  In Fig.~\ref{fig2} we show the densities for the cases of $N=5$. For larger $N$, these integrals become harder to calculate; however, the densities at $x>0$ can be obtained by mirroring the results for $x<0$ \cite{deuretz_mdist}. 

\subsection{Spin densities and initial state for $N=5$}
By taking Eq.~\ref{onebody} for the case of $N=5$, we can calculate the spin densities for the imbalanced cases of three bosons of species $\uparrow$ and two bosons of species $\downarrow$ ($N_\uparrow=3,N_\downarrow=2$) and four bosons of species $\uparrow$ and one boson of species $\downarrow$ ($N_\uparrow=4,N_\downarrow=1$). To write the separate densities for components $\uparrow$ and $\downarrow$ we must combine the spatial and spinorial contributions; the density for component $\uparrow$, for instance, is given by \cite{deuretz2}
\begin{equation}\label{spin densities}
\rho_{\uparrow}(x)=\sum_{i=1}^{N}\rho^{i}_{\uparrow}(x),
\end{equation}
where $\rho^i_{\uparrow}=m^{i}_{\uparrow}\rho^{i}(x)$ and $m^{i}_{\uparrow}$ the probability of finding a boson of species $\uparrow$ at site $i$ and $\rho^{i}(x)$ is given by Eq.~\ref{onebody}. The value of $m^{i}$ for an eingenstate is found by exact diagonalization of Hamiltonian \ref{spinchain}, where we consider $g=100$. Since the total spin projection has to be conserved, we choose the basis to be composed only by the desired states, such as $|\uparrow\uparrow\uparrow\downarrow\downarrow\rangle,...,|\downarrow\downarrow\uparrow\uparrow\uparrow\rangle$ for the $N_\uparrow=3,N_\downarrow=2$ case and $|\uparrow\uparrow\uparrow\uparrow\downarrow\rangle,...,|\downarrow\uparrow\uparrow\uparrow\uparrow\rangle$ for the $N_\uparrow=4,N_\downarrow=1$ case. 

The complete ground state wave function, including the spatial and spin eigenfunctions, must take into account the combined symmetry of these states. For instance, for a bosonic system, the ground state of the spin Hamiltonian is symmetric, which means that the spatial part of the wave function must also be symmetric to account for a totally symmetric state \cite{deuretz_mdist}. In the following sections, however, we do not take the complete wave function into account since we are dealing directly with the spin densities given by Eq.~\ref{spin densities}.

\begin{figure}[H]
\centering
\includegraphics[scale=0.43]{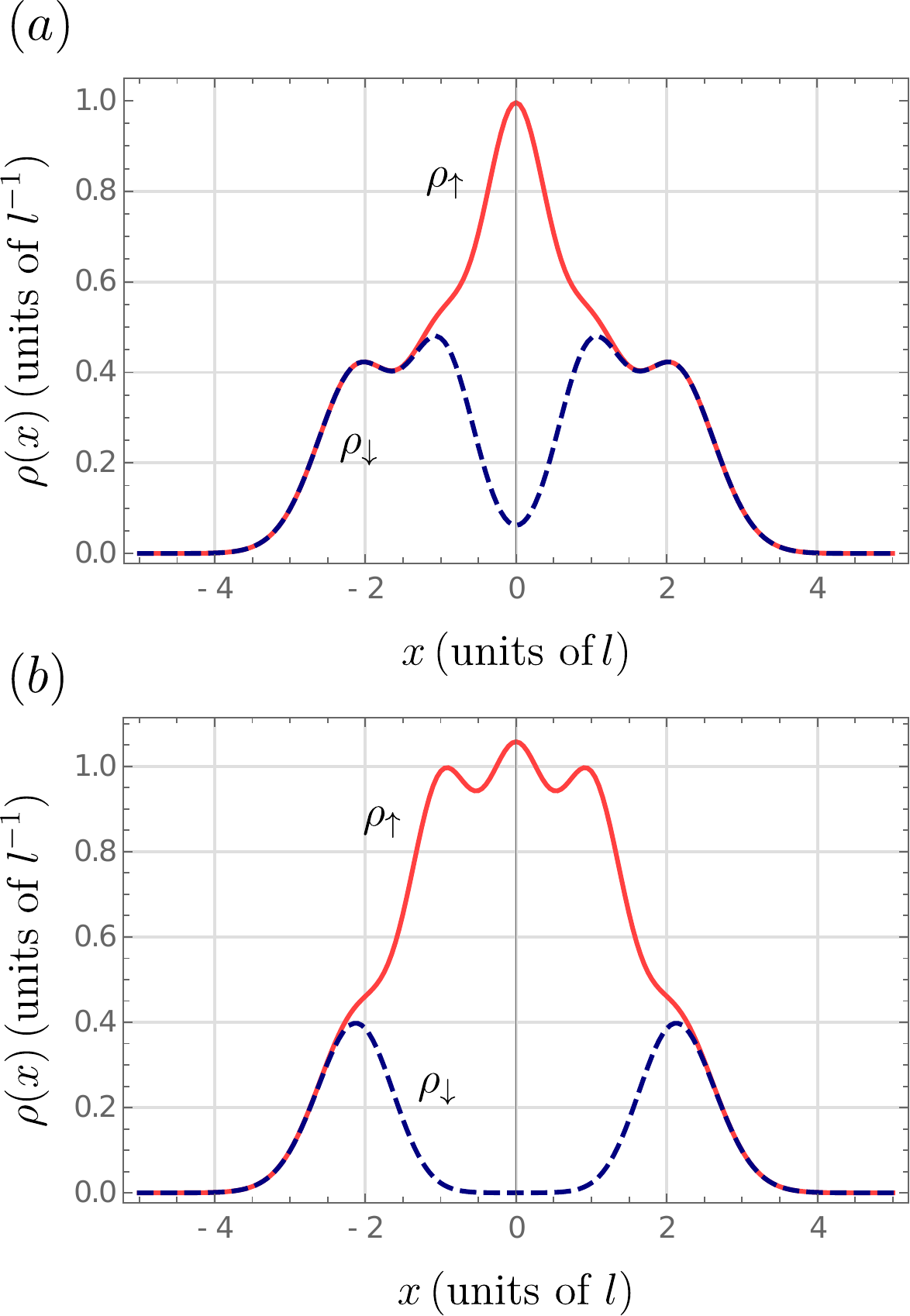}
\caption{Spin densities for the initial states, with $\kappa=0.1$, for the $(a)$ $N_\uparrow=3,N_\downarrow=2$ and $(b)$ $N_\uparrow=4,N_\downarrow=1$ cases. Solid (light red) and dashed (dark blue) curves describe the spin densities for the $\uparrow$ and $\downarrow$ components, respectively. The separation of different components in the trap indicates a FM behavior.}
\label{fig3}
\end{figure}

We now construct the initial states of the system by choosing the ground states in which the intraspecies interaction is smaller than the interspecies interaction ($\kappa=0.1$). In Fig.~\ref{fig3} we show the spin densities for the imbalanced cases of $N_\uparrow=3,N_\downarrow=2$ and $N_\uparrow=4,N_\downarrow=1$. At this point, due to the difference in the interaction strengths, the species tend to separate in the trap. The densities profiles for $\kappa<1$ show a ferromagnetic order \cite{amin1} of the Ising type, as opposed to the case where $\kappa=1$, which will be addressed next. In panel $(b)$, we see a density that is similar to that of the Bose polaron \cite{amin2}, where a strongly interacting impurity is pushed to the edges of the system.

\section{Dynamical Preparation of Magnetic States}\label{dyn}
\subsection{Imbalanced System}
We now consider the time evolution of the system for a slow increase in the intraspecies interaction parameter $\kappa$. We take $\kappa$ varying in the interval $[0.1,10.1]$. The eigenfunctions of the spin chain Hamiltonian thus evolve as
\begin{equation}
|\chi(t_f)\rangle=U(t_f,t_0)|\chi_0\rangle
\end{equation}
where $U(t_f,t_0)$ is the time evolution operator and $|\chi_0\rangle$ is the initial state. Since the hamiltonian is time dependent, we can break the time evolution in several steps
\begin{equation}
|\chi(t_f)\rangle = U(t_N,t_{N-1})...U(t_2,t_1)U(t_1,t_0)|\chi_0\rangle
\end{equation}
increasing $\Delta\kappa=10^{-5}$ and taking the Hamiltonian to be constant at each time step.

\begin{figure}[H]
\centering
\includegraphics[scale=0.43]{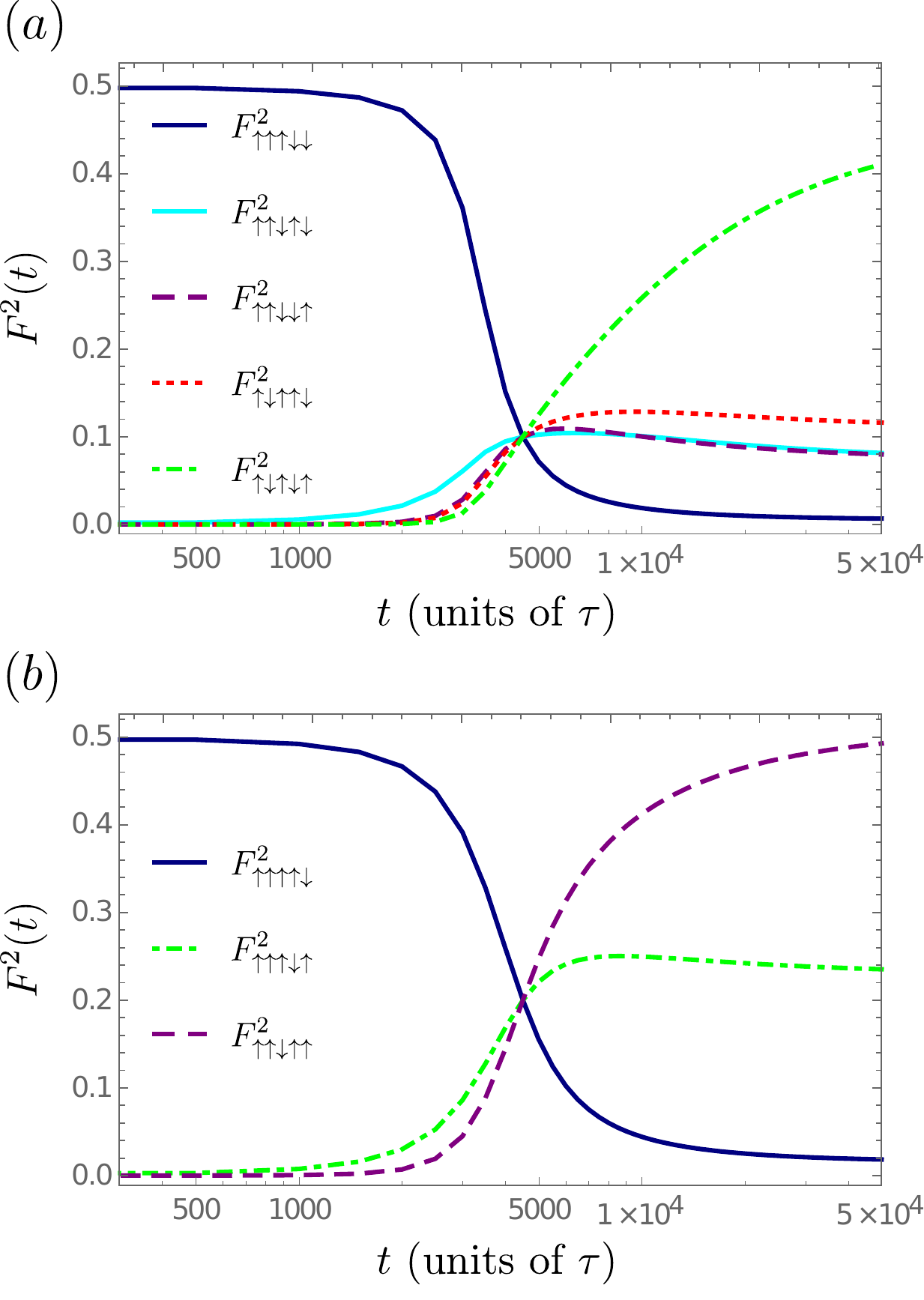}
\caption{Time evolution of the squared fidelities for $(a)$ the $N_\uparrow=3,N_\downarrow=2$ and $(b)$ the  $N_\uparrow=4,N_\downarrow=1$ cases. Identical results for symmetric states (e.g. $F^{2}_{\downarrow\downarrow\uparrow\uparrow\uparrow}=F^{2}_{\uparrow\uparrow\uparrow\downarrow\downarrow}$) are omitted. At $t=0.45\times 10^4\,[\tau]$, the system reaches the Heisenberg type FM state characterized by $\kappa=1$, where the values for all the projections are the same.}
\label{fig4}
\end{figure}

During the first steps of the time evolution ($\kappa\sim 0.1$) the change in energy at each step is larger than the spin gap $\Delta E$ between the ground state and the first excited state of Hamiltonian \ref{spinchain}. This means that, initially, the evolution of the system is not adiabatic. Therefore, the whole set of eigenvalues and eigenstates of the spin chain must be calculated for all times. The energy gap between the ground state and the first excited state of the spatial wave function, however, is given by $\hbar \omega\gg \Delta \kappa \hbar \omega$, so we can neglect the excited states of $\Phi(x_1,x_2,...,x_N)$. 

The recursion formula for the time evolution of the spin wave function is then given by
\begin{equation}
|\chi_{i+1}\rangle=\sum_{n=1}^{\nu} c^{i+1}_{n}e^{-iE^{i+1}_{n}\Delta t}|\phi^{i+1}_{n}\rangle,
\end{equation}
where $i$ denotes the time step, $E^{i+1}_{n}$ and $|\phi^{i+1}_{n}\rangle$ are the eigenvalues and eigenvectors of the Hamiltonian~\ref{spinchain} at step $i+1$, $c_{n}^{i+1}=\langle \phi^{i+1}_{n}|\chi_{i}\rangle$ and $\nu$ is the number of eigenstates (the total time evolution may be thought of as a succession of small quenches, with fixed $\Delta t=0.05\,[\tau]$). In Fig.~\ref{fig4} we show the dynamical squared fidelities $F^2_{\xi}(t)$, with $F_{\xi}(t)=|\langle \xi|\chi(t) \rangle|$, where $|\xi\rangle$ is some basis state (e.g. $|\xi\rangle=|\uparrow\uparrow\uparrow \downarrow\downarrow\rangle$ for the $N_\uparrow=3,N_\downarrow=2$ case). Since the eigenstates are composed of linear combinations of the symmetric basis states, in Fig.~\ref{fig4}, the results of the squared fidelities for states such as $|\uparrow\uparrow\uparrow\downarrow\downarrow\rangle$ and $|\downarrow\downarrow\uparrow\uparrow\uparrow\rangle$ are identical. Therefore, we choose to omit the results for the symmetric cases.

\begin{figure}[H]
\centering
\includegraphics[scale=0.54]{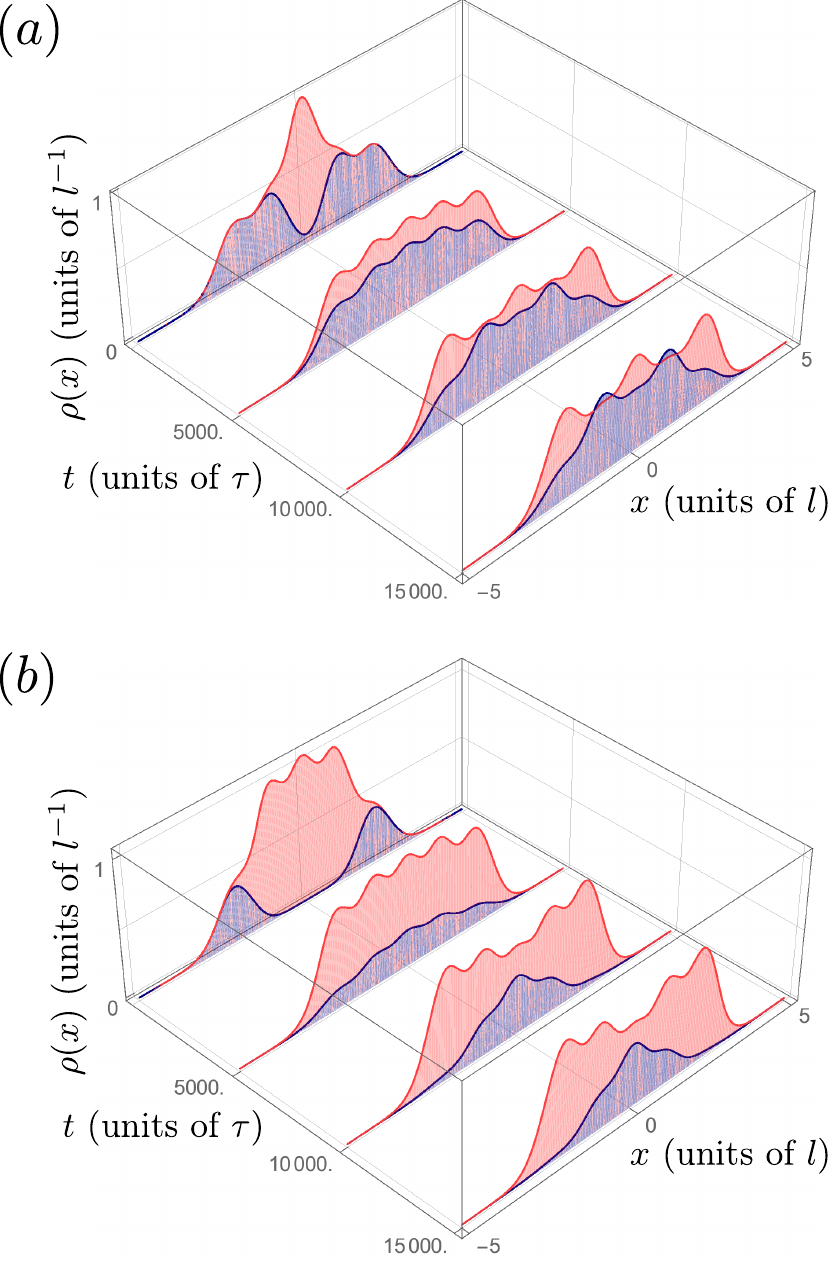}
\caption{Initial part of the time evolution ($t\leq 1.5 \times10^{4}\,[\tau]$) of the spin densities for the cases of $(a)$ $N_\uparrow=3,N_\downarrow=2$ and $(b)$ $N_\uparrow=4,N_\downarrow=1$. Light (red) and dark (blue) curves indicate the $\uparrow$ and $\downarrow$ components, respectively. Initial profiles (up to $t=0.5\times 10^{4}\,[\tau]$) indicate FM states. At around $t=10^{4}\,[\tau]$, AFM profiles start to arise.}
\label{fig5}
\end{figure}

\begin{figure}[H]
\centering
\includegraphics[scale=0.338]{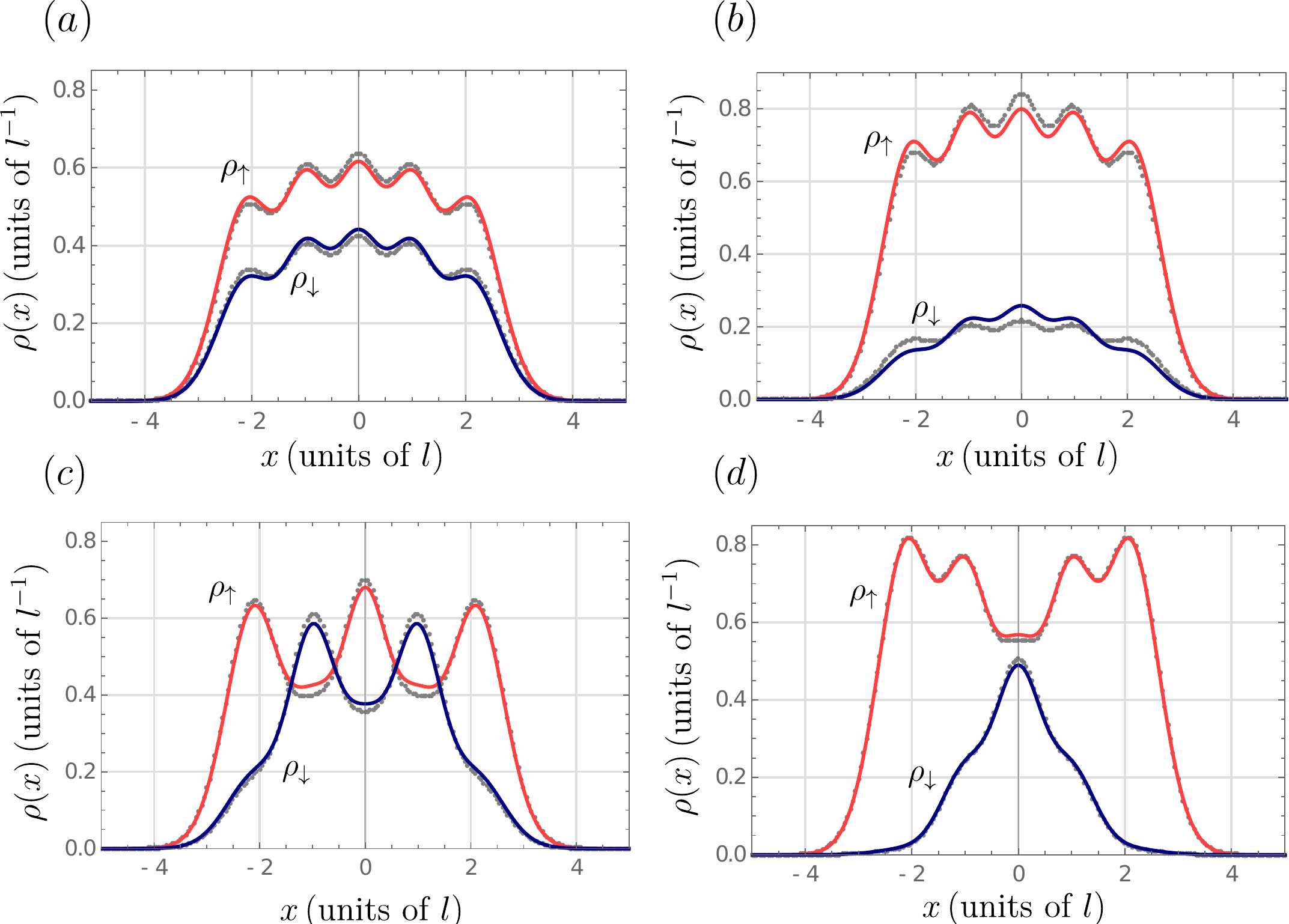}
\caption{Intermediate and final profiles for the time evolution of spin densities. Light (red) and dark (blue) curves indicate the $\uparrow$ and $\downarrow$ components, respectively. Upper panels show the profiles at $\kappa=1.1$ ($t=0.5\times 10^{4}\,[\tau]$) for the $(a)$ $N_\uparrow=3,N_\downarrow=2$ and $(b)$ $N_\uparrow=4,N_\downarrow=1$ cases. The gray dots correspond to the results obtained with $g=-100$ and $\kappa\rightarrow \infty$. The lower panels show the final profiles ($t=0.5 \times 10^{5}\,[\tau]$, $\kappa\sim 10$) for the $(c)$ $N_\uparrow=3,N_\downarrow=2$ and $(d)$ $N_\uparrow=4,N_\downarrow=1$ cases, now compared to the limiting case of $g=100$ and $\kappa \rightarrow \infty$ (gray dots).}
\label{fig6}
\end{figure}

In Fig.~\ref{fig5} we present the time evolution (up to $t= 1.5 \times10^{4}\,[\tau]$) of the spin densities for the two imbalanced cases under consideration. We see that, for $0<\kappa\leq1$, the system evolves through a FM phase. This phase is characterized first by the separation of the two components in the trap and then (around $t=0.45 \times 10^{4}\,[\tau]$ and $\kappa\sim 1$) by the typical densities of two-component bosonic systems with strong repulsive interactions \cite{hao}. For the particular case of $\kappa=1$, all the interactions between bosons are identical. The magnetic order is of the Heisenberg type with isotropic interactions, and the squared fidelities assume the same values for all basis states, as we can observe in Fig.~\ref{fig4}. In this regime, the densities show the profiles that characterize itinerant ferromagnetism (notice the distinction between the profiles in this regime and in the Ising type FM regime of $\kappa<1$). In Fig.~\ref{fig6} $(a)$ and $(b)$, we show the comparison between the slice at $t=0.5\times 10^{4}\,[\tau]$ (which corresponds to $\kappa=1.1$), and the results obtained by exactly diagonalization of Hamiltonian \ref{spinchain} with $g=-100$ and $\kappa \rightarrow \infty$. In this limit, the densities reproduce the results expected for the strongly attractive two-component fermionic gas \cite{massign}.

As the intraspecies interaction becomes stronger ($\kappa>1$) an AFM profile starts to arise. This is translated in Fig.~\ref{fig4} as the increase of the projections over the states $|\uparrow\downarrow\uparrow\downarrow\uparrow\rangle$ (green dash-dotted curve in $(a)$) and $|\uparrow\uparrow\downarrow\uparrow\uparrow\rangle$ (purple dashed curve in Fig.~\ref{fig4} $(b)$). This effect can be seen already during the first part of the time evolution ($t\geq 1.0\times10^{4}\,[\tau]$), as it is shown in Fig.~\ref{fig5}. Finally, for $\kappa\gg 1$, the AFM profiles become more pronounced (rigorously, a AFM state can only be reached for $\kappa\rightarrow \infty$). In Fig.~\ref{fig6} $(c)$ and $(d)$, we compare the final densities at $\kappa=10.1$ to the results obtained for $\kappa \rightarrow \infty$. The results in this case match the AFM states of strongly repulsive two-component fermions. It is important to point out that while the spin densities may reproduce results of fermionic systems in certain limits, this may not be true for other correlations (e.g. the momentum distribution).

The total time evolution is given by $t_f=0.5\times10^{5}\,[\tau]$. In current experimental setups, the inverse frequency $\tau$ is of the order of $100\mu$s \cite{jochim3}. This results in a total time of $5$ seconds for the process we are considering, which is a relatively long time for experiments with ultracold atoms. We point out, however, that the transition from FM to AFM-like profiles is manifested early on in this time evolution. This means that these effects could conceivably be observed in smaller time intervals. Alternatively, increasing the trap frequency could lead to smaller time scales, where the increase in the interactions would take a shorter time.

\subsection{Balanced System}
We consider now a balanced system composed of $N_\uparrow=2,N_\downarrow=2$. Once again we choose an initial state where the intraspecies interaction is smaller than the interspecies interaction ($\kappa=0.1$). 

\begin{figure}[H]
\centering
\includegraphics[scale=0.34]{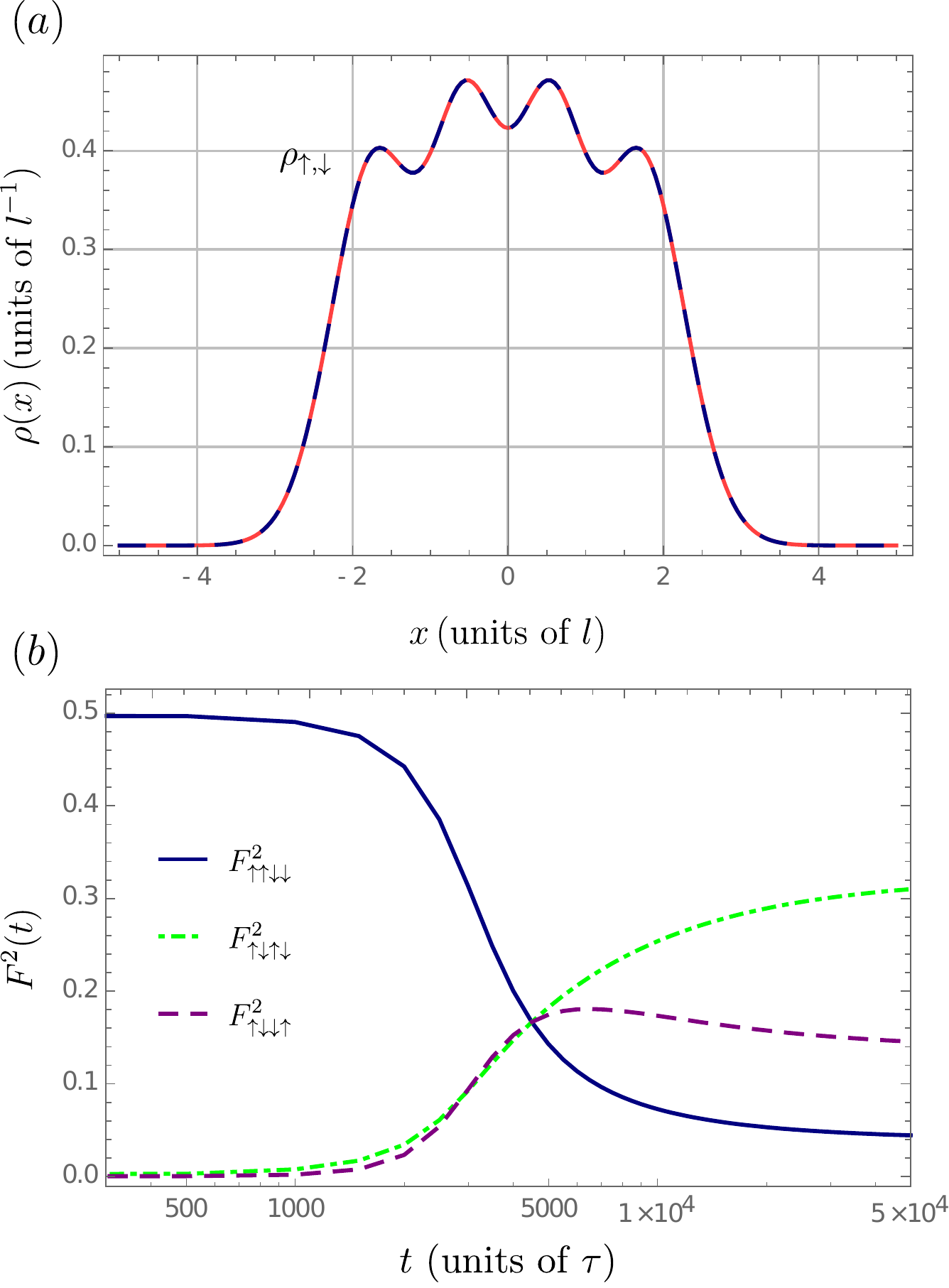}
\caption{(a) Spin densities for a balanced $N_\uparrow=2,N_\downarrow=2$ system. Solid (light red) and dashed (dark blue) curves indicate the $\uparrow$ and $\downarrow$ components, respectively. Due to the absence of imbalance, there is no change in the spin densities as $\kappa$ is varied. (b) The squared fidelities display a transition from FM to AFM states similar to those observed in the imbalanced cases.}
\label{fig7}
\end{figure}

In Fig.~\ref{fig7} (a), although a FM profile is still observed, there is no visible separation of components in the trap, due to the fact that the system is now balanced. Since the probabilities of finding spin up and down bosons at each site are always the same, the spin densities do not change in time as $\kappa$ increases. However, the squared fidelities display a behavior analogous to that of the imbalanced cases, where the AFM states become dominant as $\kappa \rightarrow \infty$. Unlike the imbalanced cases, the AFM state for $N_\uparrow=2,N_\downarrow=2$ is composed by the linear combination of $|\uparrow\downarrow\uparrow\downarrow\rangle$ and $|\downarrow\uparrow\downarrow\uparrow\rangle$.

\section{Conclusions}\label{conc}
We showed that different magnetic states can be addressed by dynamically changing the intraspecies interactions of a two-component strongly repulsive few-body bosonic gas.  Due to the strong interactions, this model can be mapped to an effective spin chain with solutions that completely determine the state of the system. By slowly increasing the interactions between the identical bosons, we are able to keep the spatial densities fixed in the ground state, while the spin eigenstates evolve in time. The spin densities then display a clear transition between FM and AFM profiles. In addition, during this evolution the system exhibits results that match the limiting cases of strong interspecies attraction or repulsion, depending only on the tuning of the parameter $\kappa$.

Future elaborations of the work presented here could be based on the study of quench dynamics in strongly interacting bosonic mixtures, now taking into account the excited states of the spatial wave function. Other interesting extensions would include the dynamics of larger ensembles of interacting bosonic gases, which could help bridge the gap between the few-body and many-body landscapes.

\begin{acknowledgments}
The authors thank Xiaoling Cui and Carlos Kuhn for useful comments. R.E.B. thanks Marcos Pérez for inspiring discussions. The following agencies
- Conselho Nacional de Desenvolvimento Científico e Tecnológico (CNPq), the Danish Council for Independent Research DFF Natural Sciences and the DFF Sapere Aude program - are gratefully acknowledged for financial support.
\end{acknowledgments}

\bibliographystyle{ieeetr}
\bibliography{biblio}

\end{document}